\documentstyle[11pt,amssymb,epsf,cite]{article}

\makeatletter \@addtoreset{equation}{section}
\makeatother

\def\ben{\begin{equation}}
\def\een{\end{equation}}

 \let\m=\mu \let\n=\nu

\let\C=\Chi

\def\nn{\nonumber} \def\bd{\begin{document}} \def\ed{\end{document}}
\def\ds{\documentstyle} \let\fr=\frac \let\bl=\bigl \let\br=\bigr
\let\Br=\Bigr \let\Bl=\Bigl
\let\bm=\bibitem
\let\na=\nabla
\let\pa=\partial \let\ov=\overline
\newcommand{\be}{\begin{equation}}
\newcommand{\ee}{\end{equation}}
\def\ba{\begin{array}}
\def\ea{\end{array}}
\def\ft#1#2{{\textstyle{\frac{\scriptstyle #1}{\scriptstyle #2}}}}
\def\fft#1#2{\frac{#1}{#2}}
\def\del{\partial}
\def\vp{\varphi}
\def\sst#1{{\scriptscriptstyle #1}}
\def\oneone{\rlap 1\mkern4mu{\rm l}}
\def\td{\tilde}
\def\wtd{\widetilde}
\def\ie{\rm i.e.\ }
\def\dalemb#1#2{{\vbox{\hrule height .#2pt
        \hbox{\vrule width.#2pt height#1pt \kern#1pt
                \vrule width.#2pt}
        \hrule height.#2pt}}}
\def\square{\mathord{\dalemb{6.8}{7}\hbox{\hskip1pt}}}
\newcommand{\ho}[1]{$\, ^{#1}$}
\newcommand{\hoch}[1]{$\, ^{#1}$}
\newcommand{\bea}{\begin{eqnarray}}
\newcommand{\eea}{\end{eqnarray}}
\newcommand{\ra}{\rightarrow}
\newcommand{\lra}{\longrightarrow}
\newcommand{\Lra}{\Leftrightarrow}
\newcommand{\ap}{\alpha^\prime}
\newcommand{\bp}{\tilde \beta^\prime}
\newcommand{\tr}{{\rm tr} }
\newcommand{\Tr}{{\rm Tr} }

\def\0{{\sst{(0)}}}
\def\1{{\sst{(1)}}}
\def\2{{\sst{(2)}}}
\def\3{{\sst{(3)}}}
\def\4{{\sst{(4)}}}
\def\5{{\sst{(5)}}}
\def\6{{\sst{(6)}}}
\def\7{{\sst{(7)}}}
\def\8{{\sst{(8)}}}
\def\n{{\sst{(n)}}}
\def\cA{{{\cal A}}}
\def\cB{{{\cal B}}}
\def\cF{{{\cal F}}}
\def\tV{\widetilde V}
\def\tW{\widetilde W}
\def\tH{\widetilde H}
\def\tE{\widetilde E}
\def\tF{\widetilde F}
\def\tA{\widetilde A}
\def\im{{{\rm i}}}
\def\tY{{{\wtd Y}}}
\def\ep{{\epsilon}}
\def\vep{{\varepsilon}}
\def\R{\rlap{\rm I}\mkern3mu{\rm R}}
\def\bD{{{\bar D}}}

\def\R{\rlap{\rm I}\mkern3mu{\rm R}}
\def\bD{{{\bar D}}}
\def\R{{{\mathbb R}}}
\def\C{{{\mathbb C}}}
\def\H{{{\mathbb H}}}
\def\CP{{{\mathbb C}{\mathbb P}}}
\def\RP{{{\mathbb R}{\mathbb P}}}
\def\Z{{{\mathbb Z}}}
\def\bA{{{\mathbb A}}}
\def\bB{{{\mathbb B}}}
\def\bC{{{\mathbb C}}}
\def\bD{{{\mathbb D}}}
\def\bE{{{\mathbb E}}}
\def\bZ{{{\mathbb Z}}}
\def\Re{{{\frak{Re}}}}
\def\Im{{{\frak{Im}}}}
\def\cosec{{\,\hbox{cosec}\,}}
\def\Gm{{\Gamma_{\!\! -}}}
\def\Gp{{\Gamma_{\!\! +}}}
\def\stan{{standard }}
\def\nonstan{{supernumerary }}

\newcommand{\auth}{W. Chen\,, 
H. L\"u and C.N. Pope}

\begin{document}

\begin{flushright}

MIFP-05-24\\
{\bf hep-th/0510081}\\
October\  2005
\end{flushright}

\vspace{10pt}

\begin{center}

{\large {\bf Mass of Rotating Black Holes in Gauged Supergravities}}

\vspace{20pt}
\auth

\vspace{20pt}

{\it George P. \&  Cynthia W. Mitchell
Institute for Fundamental Physics,\\ Texas A\& M University,
College Station, TX 77843-4242, USA}

\vspace{40pt}

\underline{ABSTRACT}
\end{center}

   The masses of several recently-constructed rotating black holes in
gauged supergravities, including the general such solution in minimal
gauged supergravity in five dimensions, have until now been calculated
only by integrating the first law of thermodynamics.  In some respects
it is more satisfactory to have a calculation of the mass that is
based directly upon the integration of a conserved quantity derived
from a symmetry principal.  In this paper, we evaluate the masses for
the newly-discovered rotating black holes using the conformal
definition of Ashtekar, Magnon and Das (AMD), and show that the
results agree with the earlier thermodynamic calculations.  We also
consider the Abbott-Deser (AD) approach, and show that this yields an
identical answer for the mass of the general rotating black hole in
five-dimensional minimal gauged supergravity.  In other cases we
encounter discrepancies when applying the AD procedure.  We attribute
these to ambiguities or pathologies of the chosen decomposition into
background AdS metric plus deviations when scalar fields are present.  
The AMD approach, involving no decomposition into background plus deviation,
is not subject to such complications.  Finally, we also calculate the
Euclidean action for the five-dimensional solution in minimal gauged 
supergravity, showing that it is consistent with the quantum statistical
relation.

{\vfill\leftline{}\vfill \vskip 10pt \footnoterule 
{\footnotesize
Research supported in part by DOE grant
DE-FG03-95ER40917.}

}

\pagebreak
\setcounter{page}{1}

\tableofcontents
\addtocontents{toc}{\protect\setcounter{tocdepth}{2}}
\newpage

\section{Introduction}

    With the discovery of the AdS/CFT correspondence, it has become
of considerable interest to study the solutions of gauged supergravities
in five and other dimensions.  Amongst the most important such solutions
are those that describe black holes.  In recent times there has been much
progress in constructing the black hole solutions of gauged supergravity,
both supersymmetric and non-supersymmetric.  For a variety of reasons,
it is of particular interest to study the solutions describing charged
rotating black holes.

   The general solution describing a non-extremal charged rotating
black hole in five-dimensional minimal gauged supergravity was
obtained recently in \cite{d5gen1}.  It is characterised by four
parameters, associated with the mass, the charge, and the two angular
momenta in two orthogonal spatial 2-planes.  It can also be viewed as
a solution in ${\cal N}=2$ gauged supergravity coupled to two vector
multiplets, in which the electric charges carried by the graviphoton
and the two additional $U(1)$ gauge fields are set equal.  A second
recently-obtained solution of the ${\cal N}=2$ theory corresponds to a
situation where the three charges are again non-zero, with two equal
and the third related to these in a fixed ratio \cite{d5gen2}.
Another solution found in \cite{d5gen2} corresponds to having only one
non-vanishing charge, and one non-vanishing rotation parameter.
Previously, solutions had been obtained in which the two rotation
parameters were set equal, and with three equal \cite{d5first} or
three unequal \cite{d5second} charges.

   In this paper, we shall investigate some aspects of the
thermodynamics of the solutions obtained in \cite{d5gen1,d5gen2}.  One
of the important quantities that one needs to know is the mass, or
energy, of the solution.  As was discussed in \cite{gibperpop}, the
energy of a rotating black hole in an asymptotically AdS background
must be calculated with considerable care, because of the
complications associated with the absence of an asymptotically flat
region at large distance, and because of the rotation.  In particular,
the absence of an asymptotically flat region means that one cannot use
the standard ADM \cite{adm} procedure for defining the energy.
Alternative approaches include that of Abbott and Deser \cite{abodes},
and the use of Komar integrals (see \cite{Magnon}, for a discussion of
this method in asymptotically AdS backgrounds). The Komar integral
definition, involving the integration of ${*dK}$ over a spatial
hypersurface at infinity, where $K=K_\mu dx^\mu$ and $K^\mu\del_\mu$
is a timelike Killing vector, suffers from the problem that the
integrand diverges at large radius.  One therefore has to make a
subtraction of a background AdS contribution in order to obtain a
finite result, and finding a way to to this unambiguously can be
somewhat problematical.  In the Abbott and Deser (AD) definition  
one also makes decomposition of
the metric in which a background AdS term is subtracted, and then integrates
certain derivatives of the difference over a spatial hypersurface at
infinity.  

    In \cite{gibperpop}, two relatively straightforward methods
were employed, for calculating the energies of the uncharged rotating
AdS black hole solutions that were found in $D=4$ dimensions \cite{carter},
$D=5$ dimensions \cite{hawhuntay} and in all dimensions $D\ge 6$
\cite{gilupapo1,gilupapo2}.  The first method involved evaluating all the
other quantities that appear in the first law of thermodynamics
\be
dE = T dS + \Omega_i dJ_i\,,\label{firstlaw}
\ee
and then integrating (\ref{firstlaw}) in order to obtain the energy $E$.
The advantage of this method is that the Hawking temperature $T$, the
entropy $S$, the angular velocities $\Omega_i$ and the angular momenta
$J_i$ are all easily calculated, with no complications associated with
divergent integrals.  The second method employed in \cite{gibperpop}
was to use the conformal mass definition of Ashtekar, Magnon and Das
\cite{ashmag,ashdas}.  This AMD definition expresses the mass in terms of
an integral
of certain components of the Weyl tensor over the spatial conformal 
boundary at infinity.
Since the metric approaches AdS asymptotically, the integrand falls
off and the integral is inherently well-defined.   It was shown in
\cite{gibperpop} that the first-law calculation and the AMD calculation
of the mass are in agreement for the uncharged rotating black 
holes.   Analogous results for the AMD masses of
the five-dimensional black holes with equal rotation parameters found in 
\cite{d5first} were obtained
in \cite{kunluc}, giving agreement with an earlier thermodynamic calculation
of the mass in \cite{madros}.  Another calculation of the masses of the
higher-dimensional uncharged rotating black holes was given in \cite{derkat},
using the Katz-Bi\v c\'ak-Lynden-Bell superpotential.  (See also
\cite{hoisma1,hoisma2,carneu} for further discussions of mass in 
asymptotically AdS spacetimes.)

    In \cite{d5gen1,d5gen2}, the energies for the charged
rotating black hole solutions were calculated using the method of
integrating the first law of thermodynamics, which reads
\be
dE = TdS + \Omega_i dJ_i + \Phi\, dQ\,,
\ee
where $\Phi$ is the electrostatic potential difference between the
horizon and infinity, and $Q$ is the conserved electric charge.
In the present paper, we shall instead calculate the energies 
using the the Ashtekar-Magnon-Das approach.  As we shall see, the results 
agree with the earlier calculations based on the integration of
the first law of thermodynamics.  Establishing this consistency is 
important, because it makes a direct connection between the mass and
the integration of conserved quantities.

   A further test of the thermodynamic properties of the black hole
solutions is provided by calculating the Euclidean action $I$, since
the partition function in a Gibbs ensemble at fixed
temperature $T$, angular velocity $\Omega_i$ and electrostatic potential
$\Phi$ should be given by
\be 
Z(T,\Omega_i,\Phi) = e^{-\beta\, \Phi_{\rm thermo}}\,,
\ee
where $\beta=1/T$ and $\Phi_{\rm thermo}$ denotes the thermodynamic
potential.  On the other hand, in the one-loop quantum gravity approximation
the partition function is given by $Z= e^{-I}$, and so one has the 
{\it Quantum Statistical Relation}, or QSR,
first proposed for quantum gravity in \cite{gibhaw}, that
\be
\Phi_{\rm thermo} \equiv E- TS - \Omega_i J_i - \Phi Q = I T\,.
\label{qsr}
\ee
In this paper, we also calculate the Euclidean action for the
general five-dimensional rotating black hole in minimal gauged supergravity,
and verify that it is indeed consistent with the quantum statistical
relation.  

   The AMD construction gives a conformal definition of a conserved
quantity $Q[K]$ associated to any asymptotic Killing field $K$ in an
asymptotically AdS spacetime \cite{ashmag,ashdas}.  We shall summarise
the AMD method in the notation of \cite{gibperpop}.  We 
assume that asymptotically, the $D$-dimensional metric satisfies the 
Einstein equations
\be
R_{\mu\nu} = -(D-1)\, l^{-2}\, g_{\mu\nu}\,,
\ee
where $l$ is the length-scale of the asymptotically AdS metric.  In canonical
AdS coordinates, the metric therefore approaches
\be
ds^2 = -(1+y^2\, l^{-2})\, dt^2 + \fft{dy^2}{1+ y^2\, l^{-2}} +
    y^2 \, d\Omega_{D-2}^2\label{asads}
\ee
at large distance $y$.   

    Consider an asymptotically AdS 
bulk spacetime $\{X,g\}$, equipped with a conformal boundary 
$\{\del X, \bar h\}$.
It admits a conformal compactification $\{\bar X, \bar g\}$ if $\bar X
=\sqcup \del X$ is the closure of $X$, and the metric $\bar g$ extends
smoothly onto $\bar X$ where $\bar g = \Omega^2\,
g$ for some function $\Omega$ with $\Omega>0$ in $X$ and $\Omega=0$ on
$\del X$, with $d\Omega\ne0$ on $\del X$.  One might, for example, take 
\be
\Omega= \fft{l}{y}\,.\label{omegay}
\ee
Since $\Omega$ is determined only
up to a factor, $\Omega\rightarrow f\, \Omega$, where the function
$f$ is non-zero on $\del X$, the metric $\bar g$ on $\bar X$ and its
restriction $\bar h=\bar g|_{\del X}$ are defined only up to a non-singular 
conformal factor.  The conformal equivalence class $\{\del \bar X,\bar h\}$ is
called the conformal boundary of $X$.  If $\bar C^\mu{}_{\nu\rho\sigma}$ is
the Weyl tensor of the conformally rescaled metric $\bar g_{\mu\nu}=
  \Omega^2\, g_{\mu\nu}$, and $\bar n_\mu \equiv \del_\mu \Omega$, then
in $D$ dimensions one defines
\be
\bar{\cal E}^\mu{}_{\nu}= l^2 \Omega^{D-3}\, \bar n^\rho \, \bar n^\sigma\,
    \bar C^\mu{}_{\rho\nu\sigma}\,.\label{cale}
\ee
This is the electric part of the Weyl tensor on the conformal boundary.
The conserved charge $Q[K]$ associated to the asymptotic Killing vector
$K$ is then given by
\be
Q[K]= \fft{l}{8\pi\, (D-3)}\, \oint_{\Sigma}\,
       \bar{\cal E}^\mu{}_{\nu}\, K^\nu\, d\bar\Sigma_\mu\,,\label{conffor}
\ee
where $d\bar\Sigma_\mu$ is the area element of the $(D-2)$-sphere section of
the conformal boundary.\footnote{The derivation of (\ref{conffor}) is
discussed in \cite{ashmag,ashdas}.  The key point is that by making use of
Bianchi identities, one can show that
\be
\bar D^\nu \bar{\cal E}_{\mu\nu} = 8 \pi (D-3)\, \bar T_{\nu\rho}\,
         \bar n^\nu\, \bar h^{\rho}{}_\mu\,,
\ee
where $\bar D_\mu$ is the covariant derivative in the conformal boundary
metric $\bar h_{\mu\nu}$, and $\bar T_{\mu\nu} = \Omega^{2-D}\, T_{\mu\nu}$,
in the limit as the boundary is approached, where $T_{\mu\nu}$ is the
energy-momentum tensor in the bulk.}  Note that the expression (\ref{conffor})
is invariant under the non-singular conformal transformations of the
boundary metric that we discuss above.  Thus, one may take for $\Omega$ 
any conformal factor that is related to (\ref{omegay}) by a non-singular 
multiplicative factor.

   In order to define the energy, one takes $K=\del/\del t$, where $t$
is the time coordinate appearing in the asymptotic form (\ref{asads})
of the metric under investigation.  The energy is then given
by\footnote{Note that some confusion in earlier literature arose when
coordinate systems that were rotating asymptotically at infinity were
used in an attempt to define the mass.  As emphasised in
\cite{gibperpop}, one should define the mass with respect to an
asymptotically-static coordinate system.  Especially, when considering
the thermodynamics of the system, it is highly advantageous to avoid
using an asymptotically rotating coordinate system whose rotation rate
depends on the parameters of the black hole \cite{gibperpop2}. (For
what appear to be largely historical reasons, rotating AdS black holes
were often presented in such rotating coordinate systems.)  Of course one could
always readjust all calculations so as to refer them to the
asymptotically-rotating frame, but describing the physics from such a
parameter-dependent rigid rotating frame is an easily avoidable and
unnecessary complication.}
\be
E = \fft{l}{8\pi\, (D-3)}\, \oint_{\Sigma} \bar{\cal E}^t{}_t\, 
    d\bar\Sigma_t\,.\label{amdeint}
\ee

   The organisation of the paper is as follows.  In section 2, we use
the AMD definition to calculate the mass of the recently-constructed
general rotating black hole solution of five-dimensional minimal
gauged supergravity, showing that it agrees with the earlier
calculation of the mass in \cite{d5gen1}, where it was obtained by
integrating the first law of thermodynamics.  We also calculate the
Euclidean action for the solution, and show that it is consistent with
the quantum statistical relation (\ref{qsr}).

   In section 3 we calculate the generalised AMD masses for
some recently-obtained rotating black-hole solutions of maximal gauged
five-dimensional supergravity, where there are three charges carried
by fields in the $U(1)^3$ abelian subgroup of the $SO(6)$ gauge group.
Again, we find that the AMD masses agree with the earlier results in 
\cite{d5gen2}, which were obtained by
integrating the first law.  

   Section 4 contains similar calculations of the AMD masses for 
the known rotating black-hole solutions in
four-dimensional and seven-dimensional gauged supergravities, and we
find agreement with the earlier calculations based on the integration
of the first law.

   In section 5, we consider the calculation of the black hole masses
using the methods of Abbott and Deser.  The section begins with a
brief summary of the AD procedure, in which we extend the standard
discussion of a pure Einstein theory with cosmological constant to
include to the case where there are matter fields, such as one has in
a gauged supergravity.  Especially, in gauged supergravity there is
usually a scalar potential rather than a pure cosmological constant.
We then use the AD procedure to calculate the mass of the general
rotating black hole in five-dimensional minimal gauged supergravity,
which was constructed in \cite{d5gen1}.  We find that the result
agrees with our AMD calculation in section 2, and also therefore with
the earlier calculation from the integration of the first law.  We
then consider the other rotating black holes in five, four and seven
dimensions.  We find in these four and five dimensional examples that
the AD procedure is rather tricky to implement, because of ambiguities
associated with the subtraction procedure when one separates the
metric into an AdS background plus deviations.  To study this more
fully, we look also at the rather simple examples of multi-charge
non-rotating black holes in five, four and seven-dimensional gauged
supergravities.  The AD procedure rather straightforwardly gives rise
to the correct masses in the cases when the charges are set equal.
However when the charges are unequal, implying that non-trivial scalar
fields are present in the solution, the complications of the
subtraction procedure again lead to difficulties in obtaining the
correct mass in an unambiguous manner, in the five and
four-dimensional cases.  We then discuss two possible correction terms
to the AD mass formula, incorporating additional contributions from the
scalar fields, and we relate these to corrections discussed previously in
the literature.

   Finally, the paper ends with conclusions in section 6.

\section{Rotating black holes in $D=5$ minimal gauged supergravity}

   For our principal example, we consider the recently-discovered general 
rotating black
holes in $D=5$ minimal gauged supergravity.  The Lagrangian for the
bosonic sector of the theory is given by
\be
{\cal L} = (R+ 12g^2)\, {*\oneone} - \ft12 {*F}\wedge F + \fft1{3\sqrt3}
      F\wedge F\wedge A\,,\label{d5lag0}
\ee
where $F=dA$, and the gauge-coupling $g$ is assumed to be positive, 
without loss of generality.  It is related to the asymptotic AdS radius
$l$ by $l=1/g$.   
The rotating black hole with two independent rotating parameters
is given by \cite{d5gen1}
\bea
ds^2 &=& -\fft{\Delta_\theta\, [(1+g^2 r^2)\rho^2 dt + 2q \nu]
\, dt}{\Xi_a\, \Xi_b \, \rho^2} + \fft{2q\, \nu\omega}{\rho^2}
+ \fft{f}{\rho^4}\Big(\fft{\Delta_\theta \, dt}{\Xi_a\Xi_b} -
\omega\Big)^2 + \fft{\rho^2 dr^2}{\Delta_r} +
\fft{\rho^2 d\theta^2}{\Delta_\theta}\nn\\
&& + \fft{r^2+a^2}{\Xi_a}\sin^2\theta d\phi^2 +
      \fft{r^2+b^2}{\Xi_b} \cos^2\theta d\psi^2\,,\label{5met}\\
A &=& \fft{\sqrt3 q}{\rho^2}\,
         \Big(\fft{\Delta_\theta\, dt}{\Xi_a\, \Xi_b}
       - \omega\Big)\,,\label{gaugepot}
\eea
where
\bea
\nu &=& b\sin^2\theta d\phi + a\cos^2\theta d\psi\,,\qquad
\omega = a\sin^2\theta \fft{d\phi}{\Xi_a} +
              b\cos^2\theta \fft{d\psi}{\Xi_b}\,,\nn\\
\Delta_r &=& \fft{(r^2+a^2)(r^2+b^2)(1+g^2 r^2) + q^2 +2ab q}{r^2} - 2m
\,,\nn\\
\Delta_\theta &=& 1 - a^2 g^2 \cos^2\theta -
b^2 g^2 \sin^2\theta\,,\quad
\rho^2 = r^2 + a^2 \cos^2\theta + b^2 \sin^2\theta\,,\nn\\
\Xi_a &=&1-a^2 g^2\,,\quad \Xi_b = 1-b^2 g^2\,,\qquad
f= 2 m \rho^2 - q^2 + 2 a b q g^2 \rho^2\,.\label{xiab}
\eea

\subsection{Conformal AMD mass}\label{amdmasssec}
 
    The metric (\ref{5met}) is written in an asymptotically non-rotating
coordinate system.  If one defines new coordinates $(y,\hat\theta)$ by
\be
\Xi_a\, y^2\, \sin^2\hat\theta= (r^2+a^2)\, \sin^2\theta\,,\qquad
\Xi_b\, y^2\, \cos^2\hat\theta= (r^2+b^2)\, \cos^2\theta\,,\label{yrrel}     
\ee
then it can be seen to approach (\ref{asads}) at large $y$.  As we discussed
earlier, we could choose to define the boundary metric of the conformal
compactification using the conformal factor $\Omega=l/y$ (\ref{omegay}).  
In practice, however, it is simpler to take
\be
\Omega= \fft{l}{r} = \fft1{g r}\,,\label{omegar}
\ee
which is related to (\ref{omegay}) by the non-singular scale factor
$f=y/r$.  With this choice, the relevant electric component 
of the Weyl tensor is given
\begin{eqnarray}
\bar{\cal E}^{t}{}_{t}  =  
\frac{1}{g^2\Omega^2}\bar{g}^{\alpha r}\bar{g}^{\beta r} \bar n_r  
       \bar n_r C^{t}{}_{\alpha t \beta} 
                         =  \frac{1}{g^4r^4 \Omega^6}
(g^{rr})^2C^{t}{}_{r t r}\,.\label{etoc}
\end{eqnarray}
We find that as $r\rightarrow\infty$, the leading order term of
$g^{rr}$ is $ g^2r^2$, while asymptotically $C^{t}{}_{r t r}$ is given by
\begin{eqnarray}
C^{t}{}_{r t r}&=&\frac{2}{g^2\Xi_a\, \Xi_b\, r^6}\Big(
3m-3a^2g^2m+b^2g^2m-a^2b^2g^4m+4abg^2q-4a^3bg^4q  \\
& &+4a^2g^2m\sin^2{\theta}-
4b^2g^2m\sin^2{\theta}+4a^3bg^4q\sin^2{\theta}-4ab^3g^4q\sin^2{\theta}
\Big) +O(r^{-8})\,.\nn
\end{eqnarray}
Thus in the limit of large $r$ we have
\begin{eqnarray}
\bar{\cal E}^{t}{}_{t}&=&
\frac{2g^4}{\Xi_a\, \Xi_b}\Big(
3m-3a^2g^2m+b^2g^2m-a^2b^2g^4m+4abg^2q-4a^3bg^4q\nn\\
& &+4a^2g^2m\sin^2{\theta}-4b^2g^2m\sin^2{\theta}+
4a^3bg^4q\sin^2{\theta}-4ab^3g^4q\sin^2{\theta}\Big)
\end{eqnarray}

     To perform the integral, we need to find the hypersurface normal to
the Killing vector field. As $r\rightarrow \infty$, the
conformally rescaled metric is
\begin{equation}
d\bar{s}^2=-\frac{\Delta_{\theta}}{\Xi_{a}\Xi_{b}}dt^2+
\frac{1}{g^2\Delta_{\theta}}d\theta^2+\frac{1}{g^2\Xi_{a}}
\sin^2{\theta}d\phi^2+\frac{1}{g^2\Xi_{b}}\cos^2{\theta}d\psi^2
\label{confmet51}
\end{equation}
The area element $d\Sigma_\mu$ is obtained from (\ref{confmet51}) as
follows.  First, we note that the 4-volume element for the boundary
metric (\ref{confmet51}) is given by
\be
\hbox{Vol} = \fft{\sin\theta\, \cos\theta}{g^3 \Xi_a \Xi_b}\,
         dt\wedge d\theta\wedge d\phi\wedge d\psi\,.
\ee
Now, we define $d\Sigma_\mu\equiv \langle \del_\m, \hbox{Vol}\rangle$,
where the angle brackets indicate that one performs the contraction
(inner product) between the vector and the form,
using the rule $\langle \del_\mu, dx^\nu\rangle
 = \delta_\mu^\nu$. Thus we shall, in particular, have
\be 
d\Sigma_t = \fft{\sin\theta\, \cos\theta}{g^3 \Xi_a \Xi_b}\,
          d\theta\wedge d\phi\wedge d\psi\,,
\ee
where ``t'' is the coordinate-frame time index.  Performing
the integration, we find
\begin{eqnarray}
E&=& \fft1{16\pi g^4 \Xi_a\Xi_b}\, \int_0^{2\pi} d\phi\int_0^{2\pi}d\psi
       \int_0^{\pi/2} d\theta \, \sin\theta\cos\theta\,  
         \bar{\cal E}^t{}_t\nn\\
 &=&\frac{m\pi(2\Xi_a+2\Xi_b-\Xi_a\Xi_b)+2\pi
 qabg^2(\Xi_a+\Xi_b)}{4\Xi_a^2\Xi_b^2}\,.\label{amdmassres}
\end{eqnarray}
This result agrees precisely with the mass obtained in \cite{d5gen1} 
by integrating the first law of thermodynamics.

\subsection{Euclidean action and the QSR}

    In order to verify that the Quantum Statistical Relation
(\ref{qsr}) is satisfied, it is necessary to calculate the Euclidean
action, namely the integral of ${\cal L}$ given by (\ref{d5lag0}),
after Euclideanisation.  This was evaluated in \cite{hawhuntay} for
the case of the neutral Kerr-AdS solutions in four and five
dimensions, and in \cite{gibperpop} for the neutral Kerr-AdS solutions
in arbitrary dimension.  It was also evaluated for the rotating 
black hole in five-dimensional minimal gauged supergravity, 
in the case where the rotation parameters are equal, in \cite{kunluc}.

    The action has to be defined with care, since
the naive integration over the volume of the Euclideanised metric
gives infinity.  To obtain a finite action, one cuts off the
integration at some large radius $r=R$, and makes an appropriate
subtraction for an AdS metric with the same boundary.  Now, as $R$ is
sent to infinity, the subtracted action converges to a finite result.
It should be noted that there is no need to include the usual
Gibbons-Hawking boundary term involving the trace of the second
fundamental form, because this is precisely removed when the AdS
subtraction is performed.

   The integration for the black hole metric is straightforward, albeit
somewhat complicated.  In particular, we integrate the radial variable
$r$ from the Euclideanised horizon at $r=r_+$ (\ie the origin of coordinates
in the Euclidean regime) to the chosen large radius $R$.   To perform the AdS 
subtraction, we can use the metric obtained by setting the mass and charge
to zero in the black-hole metric, since then the AdS metric will be 
expressed directly in an appropriately adapted coordinate system.  There
is one subtlety concerning the matching of the boundaries of the 
black-hole metric and the AdS metric at $r=R$.  Namely, one must rescale
the Euclidean time coordinate $\tau$ in the AdS metric so that the volume
its $r=R$ boundary is the same as the volume of the $r=R$ boundary of
the black-hole metric.  If $\gamma_{\mu\nu}$ denotes the metric of the
$r=R$ boundary in the black-hole metric, and $\bar\gamma_{\mu\nu}$ is
the corresponding metric AdS boundary metric obtained by setting the 
mass and charge to zero, then we must choose a rescaled Euclidean
time coordinate $\bar\tau$ for AdS such that
\be
\int \sqrt{\gamma}\, d\theta d\phi d\psi d\tau = 
\int \sqrt{\bar \gamma}\, d\theta d\phi d\psi d\bar\tau \,,
\ee
where the integration is over the boundary at $r=R$.  Thus we must define
the rescaled Euclidean time coordinate $\bar\tau$ in the AdS background 
according to
\be
\bar\tau = \tau\, \fft{\int \sqrt{\gamma}\, d\theta}{
        \int \sqrt{\bar\gamma}\, d\theta}\,.\label{barbeta}
\ee
In particular, with $\tau$ in the black-hole metric having period $\beta=
1/T$, where $T$ is the Hawking temperature, it follows that $\bar\tau$ will 
have period
\be
\bar\beta = \beta\, \fft{\int \sqrt{\gamma}\, d\theta}{
        \int \sqrt{\bar\gamma}\, d\theta}\,.
\ee
For the metric (\ref{5met}), we find that 
\be
\bar\beta= \beta\, \Big(1 - \fft{M}{g^2\, R^4}\Big)\, + {\cal O}(R^{-5})\,.
\ee

   A further subtlety concerns the lower limit of the radial integration
in the AdS subtraction.  Expressed in terms of the $y$ coordinate
appearing in (\ref{asads}), one should integrate out from $y=0$.  
Using (\ref{yrrel}), this translates into the statement that one should
integrate out from a radius $r_0$, given by
\be
r_0^2 = -\fft{a^2\Xi_b\, \sin^2\theta + b^2\, \Xi_a\, \cos^2\theta}{
           \Xi_b\, \sin^2\theta + \Xi_a\, \cos^2\theta}\,.
\ee
(The fact that this defines an imaginary $r_0$ is merely an artefact
of the coordinate system being used here.) 

   The Hawking temperature for the metric (\ref{5met}) is given by
\cite{d5gen1}
\be
T= \fft{r_+^4[(1+ g^2(2r_+^2 + a^2+b^2)] -(ab + q)^2}{2\pi\,
         r_+\, [(r_+^2+a^2)(r_+^2+b^2) + abq]}\,.
\ee
After performing the steps described above, we then find that the Euclidean
action for the black-hole metric (\ref{5met}) is given by
\be
I_5= \fft{\pi\beta}{4\Xi_a\Xi_b}
\Big[m - g^2 (r_+^2 + a^2)(r_+^2 + b^2) -
\fft{q^2 r_+^2}{(r_+^2 + a^2)(r_+^2 + b^2)+abq}\Big]\,,\label{eucact1}
\ee

   The other relevant thermodynamic quantities were evaluated in
\cite{d5gen1}, and are given by
\bea
S&=&\fft{\pi^2 [(r_+^2 +a^2)(r_+^2 + b^2) +a b q]}{2\Xi_a \Xi_b r_+}
\,,\nn\\
\Omega_a &=& \fft{a(r_+^2+ b^2)(1+g^2 r_+^2) + b q}{
               (r_+^2+a^2)(r_+^2+b^2)  + ab q}\,,\qquad
\Omega_b = \fft{b(r_+^2+ a^2)(1+g^2 r_+^2) + a q}{
               (r_+^2+a^2)(r_+^2+b^2)  + ab q}\,,\nn\\
J_a &=& \fft{\pi[2am + qb(1+a^2 g^2) ]}{4 \Xi_a^2\, \Xi_b}\,,\qquad
J_b = \fft{\pi[2bm + qa(1+b^2 g^2) ]}{4 \Xi_b^2\, \Xi_a}\,,\nn\\
\Phi &=& \fft{\sqrt3 q r_+^2}{(r_+^2 + a^2)(r_+^2 + b^2)+abq}\,,
\qquad
Q = \fft{\sqrt3\, \pi q}{4 \Xi_a\, \Xi_b}\,.
\eea
It is now straightforward to substitute these and our expression 
(\ref{eucact1}) for the Euclidean action into (\ref{qsr}), and to 
confirm that the Quantum Statistical Relation is indeed satisfied.

\section{5-dimensional Black Holes in $U(1)^3$ Gauged Supergravity}

   The Lagrangian for the relevant bosonic sector of maximal
gauged supergravity in five dimensions is given by
\be
{\cal L} = R\, {*\oneone} - \ft12 {*d\varphi_i}\wedge d\varphi_i  -
  \ft12\sum_{i=1}^3 X_i^{-2}\, {*F^i}\wedge F^i  + 4 g^2 \,
  \sum_{i=1}^3 X_i^{-1}\, {*\oneone}  + 
 F^1\wedge F^2\wedge A^3\,,\label{d5lag}
\ee
where
\be
X_1= e^{-\fft1{\sqrt6}\varphi_1 -\fft1{\sqrt2} \varphi_2}\,,\qquad
X_2= e^{-\fft1{\sqrt6}\varphi_1 +\fft1{\sqrt2} \varphi_2}\,,\qquad
X_3 = e^{\fft2{\sqrt6}\varphi_1}\,.
\ee
In the following two subsections, we shall consider two
recently-discovered rotating black-hole solutions in this theory, and
use the AMD procedure to calculate the mass for each of them.  We
find that these masses agree with those derived previously by
integration of the first law of thermodynamics.

\subsection{A 3-charge rotating black hole}\label{3chargerotsec}

    A rotating black hole solution with two independent rotation parameters
was obtained \cite{d5gen2}.  The solution has three non-vanishing charges,
with two of them set equal, and the third in a fixed ratio to the other two.
Since the solution is rather complicated, we shall not present explicitly
here, but we refer the reader to \cite{d5gen2} for all the details.

    In \cite{d5gen2}, the metric is given in a coordinate 
system that is asymptotically rotating at infinity.      
To obtain the mass of the black hole, it is necessary first to
transform to an asymptotically non-rotating frame.  Starting from the 
metric given in \cite{d5gen2}, this is achieved by making the redefinitions
$\phi'=\phi+a g^2 t$ and $\psi'=\psi+b g^2 t$.  We then take the 
conformal factor defining the conformally-compactified boundary metric 
to be given by $\Omega=1/(g\,r)$.   Following the same steps as before, 
we have $\bar{\cal E}^t{}_t$ given by (\ref{etoc}), where at large $r$
we find the component $C^{t}{}_{r t r}$ of the Weyl tensor takes the form
\begin{eqnarray}
C^{t}{}_{r t r}&=&\frac{2m}{3g^2(1-a^2g^2)(1-b^2g^2)\,\,r^6}\Big(
3(3+4s^2)+b^4g^4s^2(1+4\sin^2\theta)\\
&&+a^4g^4s^2(5-4\sin^2\theta+b^2g^2(7-8\sin^2\theta))
-a^2g^2(9+3b^2g^2(1+2s^2)\nn\\
&&+12\sin^2\theta -s^2(17-16\sin^2\theta)- b^4g^4s^2(1-8 \sin^2\theta))
\nn\\
&&+b^2g^2(3-12\sin^2\theta-s^2(1+16\sin^2\theta))\Big) + 
{\cal O}\Big(\fft1{r^7}\Big)\,.\nn
\end{eqnarray}
It follows that 
\begin{eqnarray}
\bar{\cal E}^{t}{}_{t}&=&\frac{2g^4 m}{3(1-a^2g^2)(1-b^2g^2)}
\Big(3(3+4s^2)+ b^4g^4s^2(1+4\sin^2\theta)\nn\\
&&+a^4g^4s^2(5-4\sin^2\theta+ b^2g^2(7-8\sin^2\theta))
-a^2g^2(9+3b^2g^2(1+2s^2)\nn\\
&&+ 12\sin^2\theta-s^2(17-16\sin^2\theta) -b^4g^4s^2(1-8 \sin^2\theta))
\nn\\
&&+b^2g^2(3-12\sin^2\theta-s^2(1+16\sin^2\theta))\Big)\,.
\end{eqnarray}
The metric on the conformal boundary is again given by (\ref{confmet51}).
Integrating over the hypersurface normal to the Killing vector field
$K=\del/\del t$, we then obtain the black hole mass
\be
E=\fft{m\pi}{4\Xi_a^2\Xi_b^2}\Big(2\Xi_a + 2 \Xi_b - \Xi_a\Xi_b+
(2\Xi_a^2 + 2\Xi_b^2 + 2 \Xi_a\Xi_b - \Xi_a^2 \Xi_b -
\Xi_a\Xi_b^2) s^2\Big)\,,\label{d532mass}
\ee
where $\Xi_a$ and $\Xi_b$ are defined in (\ref{xiab}). 
This is precisely the mass found in \cite{d5gen2} by integrating the
first law of thermodynamics.

   The direct calculation of the Euclidean action is rather intricate
in this example.  As we saw in the previous case of the rotating black
hole in the minimal gauged supergravity, one always needs to perform
the subtraction of a fiducial action for a pure AdS background with a
matching boundary at large distance, in order to obtain a finite
result.  In the present case the process is rather more involved,
presumably because of the presence of scalar fields in the solution.
We shall not present a direct calculation of the Euclidean action
here.  However, since it is useful for some purposes to know the
expression for the action, we shall present the result here obtained
by substitution of the thermodynamic quantities, which were derived in
\cite{d5gen2}, into the quantum statistical relation (\ref{qsr}).  We
then find that the Euclidean action is given by
\bea
I_5&=&\fft{\pi\beta}{4\Xi_a\Xi_b}\Big[
m - g^2 (r_+^2 + a^2) (r_+^2 + b^2)\\
&&-\fft{g^2 q \Big( (r_+^2 + a^2)(r_+^2 + b^2) (4r_+^2 + a^2 + b^2) +
6q r_+^2 (2 r_+^2 + a^2 + b^2) +
8 q^2 r_+^2 \Big)}{(r_+^2+a^2)(r_+^2+b^2) + 2 q r_+^2}\Big]\,.\nn
\eea

\subsection{Single-charge rotating black hole}\label{d5singlesec}

      The solution is presented in equation (23) of \cite{d5gen2}, 
and owing to its complexity, we shall not repeat it here. 
To achieve an asymptotic non-rotating frame, 
we make the coordinate redefinition $\phi'=\phi+a g^2 c w t$.
We define the conformally rescaled metric
$\bar{g}_{ab}=\Omega^2g_{ab}$ where $\Omega=1/(g r)$. With
$n=d\Omega=-1/(g r^2)\, dr$, the relevant electric Weyl tensor
component is
\begin{eqnarray}
\bar{\cal E}^{t}{}_{t}  =  
\frac{1}{g^2\Omega^2}\bar{g}^{\alpha r}
\bar{g}^{\beta r}\bar n_r \bar n_r C^{t}{}_{\alpha t \beta} 
                       =  \frac{1}{g^2\Omega^6}g^{rr}g^{rr}
(-\frac{1}{g r^2})^2C^{t}{}_{r t r}\,.
\end{eqnarray}
As $r\rightarrow\infty$, we find that the leading order behaviour 
of $C^{t}{}_{r t r}$ is given by
\be
C^{t}{}_{r t r}=\fft{2m}{3g^2 r^6 \Xi} \Big( 4\sin^2\theta\,
 (1-\Xi)( 2wc^2 - s^2\Xi)
-\Xi(s^2 - 9w c^2 + 2 s^2 \Xi\Big) + {\cal O}\Big(\fft1{r^7}\Big)\,.
\ee
It follows that
\be
\bar{\cal E}^{t}{}_{t}= \fft{2g^4m}{3\Xi} 
\Big( 4\sin^2\theta\, (1-\Xi)( 2wc^2 - s^2\Xi)
-\Xi(s^2 - 9w c^2 + 2 s^2 \Xi\Big)\,.
\ee
The metric on the conformal boundary is given by
\begin{equation}
d\bar{s}^2=-\frac{\Delta_{\theta}}{\Xi}dt^2+
\frac{1}{g^2\Delta_{\theta}}d\theta^2+\frac{1}{g^2\Xi}
\sin^2{\theta}d\phi^2+\frac{1}{g^2}\cos^2{\theta}d\psi^2
\label{confmet52}
\end{equation}
We integrate over the hypersurface normal to the Killing vector field
and obtain the black hole mass
\begin{eqnarray}
E&=&\frac{m\pi}{4\Xi^2w(\Xi-w)}[\Xi-w(2+\Xi)+w^2\Xi(1+\Xi)]\,,
\label{d511mass}
\end{eqnarray}
in agreement with the thermodynamic calculation in \cite{d5gen2}.

    The thermodynamic quantities for this black hole were obtained
in \cite{d5gen2}.  Using these, we can derive the Euclidean
action using the quantum statistical relation (\ref{qsr}).  We obtain the 
result that
\be
I_5 = \fft{\pi\beta}{4\Xi} \Big[m - g^2 r_+^2 (r_+^2 + a^2)-
\fft{m(1-w)(2g^2 (1+w) r_+^2 + 1 - w \Xi )}{w(w-\Xi)}\Big]\,.
\ee 

\subsection{3-charge black hole with equal rotation parameters}

   The solution with three independendent charges, and with the two
rotation parameters set equal, was obtained in \cite{d5second}.  We
shall work with the solution in the variables that were used in
section 3.1 of \cite{cvgilupo}.  We find that the relevant component
of the Weyl tensor has the asymptotic form
\be
C^t{}_{rtr} = \fft{2m}{g^2 r^6}\Big(3 + a^2 g^2 + 2 \sum_i s_i^2\Big)
  + {\cal O}\Big(\fft1{r^7}\Big)\,,
\ee
where $s_i\equiv \sinh\delta_i$ and the $\delta_i$ are the charge (boost) 
parameters.  Accordingly, we find that the electric component in the
conformal boundary metric is given by
\be
\bar{\cal E}^t{}_t = 2m g^4\, \Big(3 + a^2 g^2 + 2 \sum_i s_i^2\Big)\,.
\ee
From this, we find after performing the integration in (\ref{amdeint}) that
the mass is given by
\be
E= \ft14 m\pi \Big(3 + a^2 g^2 + 2 \sum_i s_i^2\Big)\,,\label{amd3chargeeq}
\ee
which precisely reproduces the result obtained in \cite{cvgilupo} by
integrating the first law of thermodynamics. 

\section{Rotating Black Holes in $D=4$ and $D=7$ Gauged Supergravities}

\subsection{$D=4$ $SO(4)$ gauged supergravity}\label{d4sec}

    The general solution for rotating black holes in 
$D=4$, ${\cal N}=4$, $SO(4)$ gauged
supergravity were obtained in \cite{d4gauge}.  These carry two charges,
associated with the gauged fields in the $U(1)\times U(1)$ Cartan subalgebra
of $SO(4)$.  First it is
convenient to rescale the azimuthal coordinate $\phi$ in \cite{d4gauge}
by a factor of
$\Xi^{-1}$, so that it has the canonical period $2\pi$.  Then to
achieve a non-rotating coordinate system at infinity, we define a
azimuthal angle $\phi'=\phi+a g^2t$.  In the new coordinate
system, the relevant electric Weyl tensor component is given by
\begin{equation}
\bar{\cal E}^{t}{}_{t}  =  \frac{1}{g^2\Omega}\bar{g}^{\alpha r}
\bar{g}^{\beta r}\bar n_r \bar n_r C^{t}{}_{\alpha t \beta}
 = \frac{1}{g^4 r^4\Omega^5}(g^{rr})^2C^{t}{}_{r t r}\,,
\end{equation}
for the conformally scaled metric $\bar{g}_{ab}=\Omega^2g_{ab}$,
where we take $\Omega=1/(g r)$.

    We find that the leading order behaviour of 
$C^{t}{}_{r t r}$ at large $r$ is given by
\begin{eqnarray}
C^{t}{}_{r t r}&=&\frac{m}{g^2(1-a^2g^2)\,\,r^5}\Big(2-2a^2
g^2+2s_1^2-2a^2g^2s_1^2+2s_2^2-2a^2g^2s_2^2\nonumber\\
&&+3a^2g^2\sin^2\theta+3a^2g^2s_1^2\sin^2\theta+3a^2g^2s_2^2\sin^2\theta
\Big) + {\cal O}\Big(\fft1{r^6}\Big) \,,
\end{eqnarray}
and hence we obtain
\begin{eqnarray}
\bar{\cal E}^{t}{}_{t}&=&\frac{g^3m}{1-a^2g^2}
\Big(2-2a^2 g^2+2s_1^2-2a^2g^2s_1^2+2s_2^2-2a^2g^2s_2^2
\nonumber\\
&&+3a^2g^2\sin^2\theta+3a^2g^2s_1^2\sin^2\theta+3a^2g^2s_2^2\sin^2\theta
\Big)\,.
\end{eqnarray}
The metric on the conformal boundary is given by
\begin{eqnarray}
ds_4^2=-\frac{\Delta_{\theta}}{\Xi}dt^2+
\frac{1}{g^2\Delta_{\theta}}d\theta^2+
\frac{\sin^2\theta}{g^2\Xi}d\phi'^2\,.
\end{eqnarray}
Integrating over the hypersurface normal to the Killing vector
field $K=\del/\del t$, \ie
\begin{equation}
\frac{1}{g^2\Xi}\int_{0}^{2\pi}d\phi'\int_{0}^{\pi}d\theta
\sin\theta\,,
\end{equation}
we obtain the black hole mass
\begin{equation}
E=\frac{m(1+s_1^2+s_2^2)}{\Xi^2}\,.\label{d4mass}
\end{equation}
This expression for the mass agrees with the one obtained in \cite{cvgilupo}
by integrating the first law of thermodynamics. 

   The thermodynamic quantities for this black hole were obtained in
\cite{cvgilupo}.  Using these, we can derive the Euclidean action using
the quantum statistical relation (\ref{qsr}); it is given by
\be
I_4=\fft{\beta}{2\Xi}\Big[M - g^2r (r^2 +a^2) -
\fft{4a^2q_1q_2}{r(r_1r_2+a^2)} - g^2a^2 (q_1 + q_2) -
\fft{g^2(r_1^2r_2^2-r^4)}{r}\Big]\Big|_{r=r_+}\,,
\ee
where $q_i=m s_i^2$.

\subsection{$D=7$ gauged supergravity}\label{d7sec}

        The gauge group of $D=7$ gauged maximal supergravity is
$SO(5)$, which has $U(1)\times U(1)$ as its Cartan subalgebra.
Rotating black holes charged under these two $U(1)$ gauge fields, with
three equal angular momenta, were constructed in \cite{d7gauge}.  In order to
achieve an asymptotic non-rotating frame, we make a coordinate
transformation $\psi'=\psi+\frac{g}{\Xi_{-}}t$ (where $\Xi_-=1 -a g$, 
as defined in \cite{d7gauge}), starting from the 
metric given in \cite{d7gauge}.  Furthermore, it is
necessary to scale the time coordinate according to $t\rightarrow \Xi\,t$ 
so that it matches with the canonical time coordinate of AdS$_7$ at infinity, 
as defined by (\ref{asads}).  Having done this, we
find that the relevant electric component of the Weyl tensor is given by
\begin{equation}
\bar{\cal E}^{t}{}_{t} = \frac{1}{g^2\Omega^4}\bar{g}^{\alpha r}
\bar{g}^{\beta r}\bar n_r \bar n_r C^{t}{}_{\alpha t \beta}
=\frac{1}{g^2\Omega^8}g^{rr}g^{rr}(-\frac{1}{g\,r^2})^2C^{t}{}_{r t r}
\end{equation}
in the conformally rescaled metric $\bar{g}_{ab}=\Omega^2g_{ab}$
where $\Omega=1/(g\,r)$, with $n=d\Omega=-\frac{l}{r^2}dr$. 
Asymptotically as $r\rightarrow\infty$, the leading order behaviour
of $g^{rr}$ is $g^{rr}\sim g^2r^2$, and that for $C^{t}{}_{r t r}$ is
\begin{eqnarray}
C^{t}{}_{r t r}&=&\frac{m}{g^2(1-a^2g^2)\,\,r^8}
\Big(12(-1+4a^2g^2+4a^3g^3+a^4g^4)\nn\\
&&-2c_1c_2a^2g^2(-30-16a g+51a^2g^2+64a^3g^3+21a^4g^4)\nn\\
&&+(c_1^2+c_2^2)(16-52a^2g^2-40a^3g^3+45a^4g^4+64a^5g^5+21a^6g^6)\Big)
\,.
\end{eqnarray}
Therefore, we have
\begin{eqnarray}
\bar{\cal E}^{t}{}_{t}&=&\frac{g^6m}{1-a^2g^2}
\Big(12(-1+4a^2g^2+4a^3g^3+a^4g^4)\nn\\
&&-2c_1c_2a^2g^2(-30-16a
g+51a^2g^2+64a^3g^3+21a^4g^4)\nonumber\\
&&+(c_1^2+c_2^2)(16-52a^2g^2-40a^3g^3+45a^4g^4+64a^5g^5+21a^6g^6)
\Big)\,.\nn
\end{eqnarray}
The metric on the conformal boundary is given by
\begin{eqnarray}
ds_7^2=-\frac{1}{\Xi}dt^2+\frac{1}{g^2\Xi}d\Omega_5^2
\end{eqnarray}
where $d\Omega_5^2$ is the standard metric on the unit 5-sphere. Integrating
over the hypersurface normal to the Killing vector field $K=\del/\del t$, 
we obtain the black hole mass
\begin{eqnarray}
E&=&\frac{m\pi^2}{32\Xi^4}\Big[(12(-1+4a^2g^2+4a^3g^3+a^4g^4)\nn\\
&&\qquad -2c_1c_2a^2g^2(-30-16a
g+51a^2g^2+64a^3g^3+21a^4g^4)\nonumber\\
&&\qquad +(c_1^2+c_2^2)(16-52a^2g^2-40a^3g^3+45a^4g^4+
         64a^5g^5+21a^6g^6)\Big]\,.\label{d7mass}
\end{eqnarray}
This agrees with the mass that was calculated in \cite{cvgilupo} by 
integrating the first law of thermodynamics.   

   Again, we can present an expression for the Euclidean action for
this rotating seven-dimensional black hole, by substitution of the
thermodynamic quantities, which were calculated in \cite{cvgilupo},
into the quantum statistical relation (\ref{qsr}).  The expression we
obtain is rather complicated in the general case when the two charges
are unequal, and so here we shall just present the result when the
charges are set equal. We then define a charge parameter $q$ by 
setting $c_1=c_2=\sqrt{1+ q/m}$, and find that the Euclidean action is
given by
\be
I = \fft{\beta\, \pi^2}{8\Xi^3}\, \Big(m - g^2 R_+^6 - g^2\, q\, (4R_+^2-a^2)
  - \fft{4g q^2[g R_+^4 + a^2 g (1+ag) R_+^2 + 2gq - a^3(1+ag)^2]}{
      R_+^6 + 2q R_+^2 - 2a^2(1+ag)}\Big)\,,
\ee
where $R_+$ is the radius of the outer horizon.

\section{Abbott-Deser Mass for the Rotating Black Holes}

\subsection{The Abbott-Deser mass in gauged supergravity}
\label{adsec}

   In the Abbott-Deser AD construction \cite{abodes}, one splits the
asymptotically-AdS metric $g_{\mu\nu}$ in the form 
\be
g_{\mu\nu}= \bar g_{\mu\nu} + h_{\mu\nu}\,,\label{hdef} 
\ee
where $\bar g_{\mu\nu}$ is the AdS metric.
We shall summarise the procedure here, including the extension needed for
qconsidering asymptotically-AdS spacetimes
as solutions of gauged supergravities, where there is a scalar potential
with a stationary point rather than a pure cosmological constant.  

   Consider a $D$-dimensional theory whose bosonic Lagrangian is
\be
{\cal L}= \sqrt{-g}\, [R - V(\phi)] + {\cal L}_{\rm kin}\,,
\ee
where $\phi$ represents the scalar fields, with potential $V(\phi)$, 
and ${\cal L}_{\rm kin}$ denotes the kinetic terms for the scalars and
the other matter fields in the theory. 
We assume that $V(\phi)$ has a stationary point at $\phi=0$, and that
there exists a pure AdS background solution with $g_{\mu\nu}=
\bar g_{\mu\nu}$ and $\phi=0$, with all other fields vanishing too, where
\be
\bar R_{\mu\nu} - \ft12 \bar R\, \bar g_{\mu\nu} + \ft12 V(0)\, 
                 \bar g_{\mu\nu}=0\,.\label{bareq}
\ee
The extension of the AD prescription involves taking the full Einstein 
equation,
\be
R_{\mu\nu} - \ft12 R\, g_{\mu\nu} +\ft12 V(\phi)\, g_{\mu\nu} 
    = T^{\rm matter}_{\mu\nu}\,,\label{einst}
\ee
where $T^{\rm matter}_{\mu\nu}$ represents the energy-momentum tensor
for the other matter fields and the remaining contribution from the scalars.
Substituting $g_{\mu\nu} =\bar g_{\mu\nu} + h_{\mu\nu}$ into (\ref{einst}),
one 
then separates the terms linear in $h_{\mu\nu}$ from the remainder,
which will acquire the interpretation of an effective energy-momentum
tensor for gravity plus the other fields.  The appropriate integral involving
the effective energy-momentum tensor will then yield the mass. 
Collecting the terms linear in $h_{\mu\nu}$ on the left-hand side, 
we shall have
\be
R_L^{\mu\nu} - \ft12 R_L\, \bar g^{\mu\nu} + \fft1{D-2}\, V(0)\, h^{\mu\nu} 
 = \fft1{8\pi\sqrt{-\bar g}}\,T^{\mu\nu}\,,\label{rlint}
\ee
where $R_L^{\mu\nu}$ and $R_L$ denote the linearised Ricci tensor and
Ricci scalar, and $T^{\mu\nu}$ is the effective energy-momentum tensor 
density, including the contribution from gravity as well as from the matter
fields.  Note that the contribution from the scalar fields on the 
right-hand side is of the form
\be
\fft1{\sqrt{-\bar g}}\, T^{\mu\nu}_{\rm scal}= 
 \fft1{\sqrt{-\bar g}}T^{\mu\nu}_{\rm kinetic} - \ft12 V(\phi)\, 
   g_{\mu\nu} + \ft12 V(0)\, \bar g_{\mu\nu}\,,\label{scalart}
\ee
since the effective cosmological constant $\ft12 V(0)$ in the background 
AdS metric $\bar g_{\mu\nu}$ has been included on the left-hand side of 
(\ref{rlint}).

    As in \cite{abodes}, one defines
\bea
H^{\mu\nu} &=& h^{\mu\nu} - \ft12 \bar g^{\mu\nu}\, h^\rho{}_\rho\,,\nn\\
K^{\mu\nu\rho\sigma} &=& \ft12( \bar g^{\mu\sigma}\, H^{\rho\nu} +
   \bar g^{\rho\nu}\, H^{\mu\sigma} - \bar g^{\mu\rho}\, H^{\nu\sigma}
         - \bar g^{\nu\sigma}\, H^{\mu\rho})\,,
\eea
where here, and in what follows, 
all indices are raised and lowered using the background AdS metric 
$\bar g_{\mu\nu}$.  It follows that the left-hand side in (\ref{rlint})
is given by
\bea
&&R_L^{\mu\nu} - \ft12 R_L\, \bar g^{\mu\nu} + \fft1{D-2}\, V(0)\, h^{\mu\nu} 
\nn\\
&&= \ft12(\bar\nabla_\lambda\bar\nabla^\mu H^{\lambda\nu} + 
         \bar\nabla_\lambda\bar\nabla^\nu H^{\lambda\mu}
   -\bar{\square} H^{\mu\nu} - \bar\nabla_\alpha\bar\nabla_\beta
       H^{\alpha\beta}\, \bar g^{\mu\nu}) - \fft{V(0)}{D-2}\, 
                H^{\mu\nu}\,,\nn\\
&&= \bar\nabla_\alpha\bar\nabla_\beta K^{\mu\alpha\nu\beta} + 
   \ft12 \bar R^\mu{}_{\alpha\beta}{}^\nu\, H^{\alpha\beta} 
            -\fft{V(0)}{2(D-2)}\, H^{\mu\nu}\,.
\eea
A straightforward calculation shows that the divergence of this
quantity, with respect to the background covariant derivative 
$\bar\nabla_\mu$, vanishes identically, upon the use of the background
Einstein equation (\ref{bareq}).

   Taking $\bar\xi^\mu\del_\mu = \del/\del t$ as the canonically-normalised 
timelike Killing vector, the generalised AD mass
is then given by 
\bea
E &=&
-\fft1{8\pi}\, \oint d^{D-1}x\, T^{t\nu}\, \bar\xi_\nu\,,\nn\\
 &=&  \fft1{8\pi}\, \oint dS_i {\cal M}^i\,,\nn\\
 {\cal M}^i&=& -\sqrt{-\bar g}\, \Big[ 
\bar\xi_\nu\, \bar\nabla_\mu K^{ti\nu \mu} - 
           K^{tj\nu i}\, \bar\nabla_j \bar\xi_\nu\Big]\,,\label{admass}
\eea
where $dS_i$ is the area element of the
spatial surface at large radius, and the $t$ superscript denotes a
coordinate index in the time direction.\footnote{The original
four-dimensional Abbott-Deser construction was generalised to
arbitrary spacetime dimensions in \cite{destek}. We adopt the
normalisation given in (\ref{admass}) for all dimensions (rather than
the one chosen in \cite{destek}), since this accords with the
conventions for the definition of mass appearing in most of the
earlier literature, and, in particular, the definitions in
\cite{gibperpop}.}  Note that the index $t$ denotes the time
coordinate index, Greek indices range over all spacetime directions,
and Latin indices range over the spatial directions.  Eventually, one
sends the radius to infinity.  It should be emphasised that one must
choose a coordinate frame with respect to which the deviation
$h_{\mu\nu}$ of the full metric $g_{\mu\nu}$ from the background AdS
metric $\bar g_{\mu\nu}$ tends to zero appropriately at infinity.
 
   The AD definition was used recently
in \cite{dekate} to calculate the masses of the higher-dimensional 
neutral Kerr-AdS black holes constructed in \cite{gilupapo1,gilupapo2}.

\subsection{Rotating black hole in five-dimensional minimal gauged 
          supergravity}\label{d5minsec}

   We first apply the AD procedure described above to the case of the
general rotating black hole in five-dimensional minimal gauged supergravity.
Note that there are no scalar fields in the minimal gauged supergravity
theory, and one has just a cosmological constant, as give in (\ref{d5lag0}).
The calculation is a purely mechanical one, although of such a complexity that
it is most easily carried out
with the aid of a computer.  We find that at large distance, the relevant
integrand in the expression (\ref{admass}) for the AD mass takes the
form
\bea
&&\sqrt{-\bar g} \, 
(\bar\xi_\nu\bar\nabla_\mu K^{tr\nu \mu} - K^{tj\nu r}\, \bar\nabla_j 
  \bar\xi_\nu) \\
&&=
\!\!
-\fft{\sin\theta\cos\theta}{\Xi_a^2\Xi_b^2}\, [\Xi_b(3m+a^2 g^2 m + 4ab q g^2)
-4(a^2-b^2)g^2(m+abqg^2)\cos^2\theta ] +
   {\cal O}\Big(\fft1{r}\Big)\,.\nn
\eea
After integration over the angular coordinates, it follows from (\ref{admass})
that the mass is given by
\begin{eqnarray}
E&=& -\fft1{8\pi}\, \int_0^{2\pi} d\phi\int_0^{2\pi}d\psi
       \int_0^{\pi/2} d\theta \,  \sqrt{-\bar g} \, 
(\bar\xi_\nu\bar\nabla_\mu K^{tr\nu \mu} - K^{tj\nu r}\, \bar\nabla_j 
  \bar\xi_\nu)              \nn\\
 &=&\frac{\pi m(2\Xi_a+2\Xi_b-\Xi_a\Xi_b)+2\pi
 qabg^2(\Xi_a+\Xi_b)}{4\Xi_a^2\Xi_b^2}\,.\label{admassres}
\end{eqnarray}
This result agrees precisely with the mass obtained in \cite{d5gen1} 
by integrating the first law of thermodynamics, and that we obtained in
section \ref{amdmasssec} by applying the AMD procedure.

\subsection{AD masses for the other rotating black holes}

   In this subsection, we apply the AD procedure to the other examples
of rotating black holes discussed previously in this paper.  We find that
in the case of the other five-dimensional black holes, and the four-dimensional
black holes, the answers do not agree with the AMD and thermodynamic results.
By contrast, the mass of the seven-dimensional black hole obtained by the
AD procedure does agree with the AMD and thermodynamic calculations.

\subsubsection{3-charge rotating black hole}\label{3chargebhsec}

   First, we consider the five-dimensional 3-charge rotating black
hole discussed in section \ref{3chargerotsec}, with two equal charges
and a third in a fixed ratio to these \cite{d5gen2}.  This solution
involves non-trivial scalar fields.  In the coordinate frame used in
\cite{d5gen2}, the $h_{\mu\nu}$ components already fall off at large $r$.  
We find that ${\cal M}^i$ in (\ref{admass}) is given by
\bea
{\cal M}^r &=& 
\fft{m\sin\theta\, \cos\theta}{3\Xi_a^2 \Xi_b^2}
\Big( \Xi_b (3 (4-\Xi_a) +
   (6\Xi_a + 12 \Xi_b + \Xi_a^2 - 7 \Xi_a \Xi_b) s^2)  \\
&& + 4 \cos^2\theta (\Xi_a - \Xi_b) (3 + s^2 (3 \Xi_a+3\Xi_b-
2\Xi_a\Xi_b)\Big) -
\fft{4g^2 m^2 s^4\sin\theta\, \cos\theta}{3\Xi_a\Xi_b} + 
  {\cal O}\Big(\fft1{r}\Big)\,,\nn
\eea
which leads to the mass
\be
E'=E  - \fft{g^2 m^2 \pi s^4}{3\Xi_a\Xi_b}\,,\label{ep1}
\ee
where $E$ is the mass found in \cite{d5gen2} by integrating the first law,
and which we reproduced in (\ref{d532mass}) by using the AMD procedure.

\subsubsection{Single-charge rotating black hole}\label{1chargebhsec}

    Next, we consider the single-charge rotating black hole in five
dimensional, whose AMD mass we calculated in \ref{d5singlesec}.  
 To apply the AD procedure in this case, it is necessary first
to make the coordinate transformation
\be
r\rightarrow r (1-\frac{m s^2}{3 r^2})\,,
\ee
to ensure that $h_{\mu\nu}$ falls off at large distance.  Then we find that
\bea
{\cal M}^r&=&\frac{m \sin\theta \,\cos{\theta}}{3\Xi^2}\left [4
\cos^2{\theta}(\Xi - 1)(\Xi s^2 - 3(1 + s^2)w) - \Xi(s^2(1 + 2\Xi)
 - 9w(1 + s^2))\right ]\nn\\
&& - \frac{4 g^2 m^2 s^4 \sin\theta\, 
      \cos{\theta}}{3 \Xi} + {\cal O}\Big(\fft1{r}\Big)\,.
\eea
Thus performing the integral as in (\ref{admass}), we obtain the mass
\be
E'=E - \fft{g^2 m^2 \pi s^4}{3\Xi}\,,\label{ep2}
\ee
where $E$ is the mass obtained in \cite{d5gen2} by integrating the 
first law,  and reproduced in (\ref{d511mass}) by applying the AMD procedure.

\subsubsection{3-charge black hole with equal rotation parameters}
\label{3chargeeqsec}

    For the solution obtained in \cite{d5second}, which has three 
unequal charges and the two rotation parameters set equal, we use the
notation and conventions of section 3.1 of \cite{cvgilupo}.  First, 
it is necessary to redefine the radial coordinate $r$ according to
\be
r \longrightarrow r - \fft{m}{3r}\, \sum_i s_i^2\,,
\ee
in order that $h_{\mu\nu}$ in the decomposition (\ref{hdef}) fall off at
infinity.  Then, we find that ${\cal M}^r$, defined by (\ref{admass}), is 
given by
\be
{\cal M}^r = \ft18 m\sin\theta \Big[ 3 + a^2 g^2 + 2\sum_i s_i^2 -
   \ft43 m g^2\Big( \sum_i s_i^4 - \sum_{i<j} s_i^2 s_j^2\Big)\Big] + 
{\cal O}\Big(\fft1{r}\Big)\,.
\ee
After performing the surface integration as in (\ref{admass}), we obtain 
the expresssion 
\be
E' = E - \ft13 \pi m^2 g^2 \Big(\sum_i s_i^4 - \sum_{i<j} s_i^2 s_j^2\Big)
\label{ep3}
\ee
for the AD mass, where $E$ is the mass obtained in \cite{cvgilupo} by 
integrating the first law, and that we reproduced in this paper
using the AMD procedure in (\ref{amd3chargeeq}).

\subsubsection{Four-dimensional black hole}\label{d4bhsec}

      To apply the AD procedure to calculate the mass for the four-dimensional 
gauged supergravity black holes discussed in section \ref{d4sec},  
we first need to make the coordinate transformation
\be
r\rightarrow r - m (s_1^2 + s_2^2) + \fft{m^2 (s_1^2-s_2^2)^2}{2r}\,,
\ee
in order to ensure that $h_{\mu\nu}$ falls off at large distance. We then 
find
\be
{\cal M}^r
= \fft{m^2 g^2 (s_1^2-s_2^2)^2r\sin\theta }{1-a^2g^2} +
\fft{m (1 + s_1^2 + s_2^2) [-2 + a^2 g^2 (-1 + 3\cos^2\theta)]\sin\theta}{
(1-a^2 g^2)^2} + {\cal O}\Big(\fft1{r}\Big)\,,\label{ep4}
\ee
which actually diverges as $r$ is sent to infinity.

\subsubsection{Seven-dimensional black hole}\label{d7admasssec}

   The implementation of the AD procedure for calculating the mass of
the seven-dimensional gauged supergravity black holes that we
discussed in section \ref{d7sec} is rather straightforward.  In the
coordinate frame we are using, the components $h_{\mu\nu}$ already
tend to zero at large distance.  We then find that the AD mass
calculated from (\ref{admass}) agrees precisely with the expression
(\ref{d7mass}) which was obtained in \cite{cvgilupo}) by integrating
the first law, and that we reproduced in this paper by applying the
AMD procedure.  (We have also checked the AD calculation of the mass for
the non-rotating black hole in six-dimensional gauged supergravity found in
\cite{d6romans}, and found that it agrees with the mass calculated using the
AMD method.)

   In the following subsection, we shall discuss in more detail the
problems we encountered above in calculating the AD masses of the five
and four-dimensional black hole examples.  Before moving on to this,
it is perhaps worth pointing out that in the four cases where we have
encountered difficulties in determining the mass by the AD procedure,
summarised by equations (\ref{ep1}), (\ref{ep2}), (\ref{ep3}) and
(\ref{ep4}), one gets the correct answer if follows the ``rule of
thumb'' of retaining only the terms linear in the mass or charge, and
discarding terms of higher order in the mass or charge.

\subsection{Subtleties in the AD procedure}

    It has been remarked upon previously in the literature that the
background-subtraction prescription inherent in the AD definition of
the mass can lead to ambiguities associated with coordinate 
reparameterisations of the metric and the background.  (See, for example,
\cite{ashdas,liusab}.)  On the other hand, it has been applied
successfully to calculate the masses of the rotating AdS black holes
in arbitrary dimension \cite{dekate}, and we have applied it
successfully in this paper in the case of the five-dimensional charged
rotating black holes of five-dimensional minimal gauged supergravity,
and charged solutions of seven-dimensional gauged supergravity.  All
the cases that we tried where it failed involve solutions with
non-trivial scalar fields, and as we shall discuss below, these scalar
fields can have a quite significant contribution in the calculation of
the energy. Although the seven-dimensional black holes, for which we
obtained the correct mass by the AD approach, also involve non-trivial
scalars, we find that in this case the scalars make a less significant
contribution to the energy, in a way that we shall elaborate on below.

   It might be natural to suppose that the difficulties we have
encountered are ultimately related to some ambiguities in the
decomposition of the metric into AdS background plus deviations, and
that these ambiguities become more acute in the cases where the scalar
fields play a r\^ole.  
    
\subsubsection{Scalar fields and the AD mass formula}
\label{3chargesec}

   Although we successfully applied the AD procedure above to calculate
the mass of the rotating black hole in five-dimensional minimal 
gauged supergravity, we find that in more complicated situations
we encounter problems in extracting results by using the AD method.  In
fact the difficulties can already be illustrated if we consider the
example of 3-charge non-rotating black holes in five-dimensional 
maximal gauged supergravity, for which the relevant bosonic Lagrangian is
given by (\ref{d5lag}).

    The 3-charge non-rotating black hole in five-dimensional gauged
supergravity is given by \cite{becvsa}
\bea
ds^2 &=& - (H_1 H_2 H_3)^{-2/3}\, f\, dt^2 + (H_1 H_2 H_3)^{1/3}\, \Big(
   \fft{dr^2}{f} + r^2 d\Omega_3^2\Big)\,,\nn\\
X_i &=& H_i^{-1}\, (H_1 H_2 H_3)^{1/3}\,,\qquad
  A^i= (1- H_i^{-1})\, \coth\delta_i\, dt\,,\label{3chargesol}
\eea
where
\be
f = 1 - \fft{2m}{r^2} + g^2 r^2\, H_1 H_2 H_3\,,\qquad
    H_i = 1 + \fft{2m\, \sinh^2\delta_i}{r^2}\,.
\ee
The mass of the black hole is given by \cite{becvsa}
\be
E = \fft{\pi\, m}{4}\, \Big(3  + 2 \sum_i s_i^2\Big)\,,\label{3chargemass}
\ee
where $s_i\equiv \sinh\delta_i$. 

     In the AD calculation of the mass, the background AdS metric satisfying
(\ref{bareq}) is most easily obtained by setting the mass parameter $m$
and charge parameters $\delta_i$ to zero in (\ref{3chargesol}).  However,
in the coordinate frame used in (\ref{3chargesol}), one finds that the
components of $h_{\mu\nu}$ defined by (\ref{hdef}) 
do not fall off at large $r$.  This can be 
remedied by performing the coordinate transformation $r\rightarrow
 \rho$, where
\be
r^2 = \rho^2 -\ft23 m\, (s_1^2 + s_2^2 + s_3^2)\,.
\ee
Substituting into (\ref{admass}),
we find that at large $\rho$, the integral gives the expression
\bea
E'&=&-\fft1{8\pi}\, \oint d\Sigma_i\,\sqrt{-\bar g}\,\Big[ 
(\bar\xi_\nu \bar\nabla_\mu K^{ti\nu \mu} - K^{tj\nu i}\, \bar\nabla_j 
\bar\xi_\nu \Big]\nn\\
&=&  \fft{\pi\, m}{4}\, \Big(3  + 2 \sum_i s_i^2\Big) -\ft13g^2 m^2(
s_1^4 + s_2^4 + s_3^4 - s_1^2 s_2^2 -s_2^2 s_3^2 - s_3^2 s_1^2)\,.
\eea
This disagrees with the standard result (\ref{3chargemass}), unless one sets
the three charges equal.

    The disagreement appears to be associated with the presence of the 
non-trivial scalar fields.  In fact, if we calculate the contribution of
potential energy term in (\ref{scalart}) for this solution, we find that
this contributes 
\be
-\fft{1}{8\pi \sqrt{-\bar g}}\, T^{t\nu}_{\rm pot}\, \bar\xi_\nu  = 
\fft{g^2 m^2}{48\pi\, \rho}\,  
   \sum_{i<j} (s_i^2-s_j^2)^2 \, \sin\theta\cos\theta 
     + {\cal O}(1/\rho^3)
\ee
to the energy density.  When integrated over the spatial 4-volume, this
would give rise to a logarithmic divergence at large distance.  In fact the
scalar kinetic terms contribute an equal and opposite divergence,
\be
-\fft{1}{8\pi \sqrt{-\bar g}}\, T^{t\nu}_{\rm kinetic}\, \bar\xi_\nu  
  =- \fft{g^2 m^2}{48\pi\, \rho}\,  \sum_{i<j} (s_i^2-s_j^2)^2\, 
       \sin\theta\cos\theta +  {\cal O}(1/\rho^3)
\ee
(Of course it is the $(\del\phi)^2\, g^{\mu\nu}$ term, and not the
  $\del^\mu\phi\, \del^\nu\phi$ term in the scalar energy-momentum 
tensor that contributes here, since the solution is time-independent.)
Although the scalar energy density therefore integrates to a finite total, 
it is possibly significant that the potential and kinetic energies are
separately divergent.  

   By contrast, if we calculate the analogous scalar energy contributions 
in the case of 2-charge black holes in seven-dimensional gauged supergravity
(see, for example, \cite{10auth} for details of these solutions), we find
that both the potential and kinetic energy densities integrate to give
separately finite energy contributions.  Significantly, we find in this
case that (as also for the rotating seven-dimensional black holes discussed
in section \ref{d7admasssec}) the AD calculation of the mass agrees with the
thermodynamic calculation and the AMD calculation.

   The situation is even more striking in four dimensions.  Let us
consider the non-rotating 4-charge black holes of maximal gauged
supergravity \cite{dufliu,sab}, in the notation given in \cite{10auth}
(but with the gauge coupling rescaled according to $g\rightarrow g/2$,
so that $1/g=l$, the AdS radius).  Again, we perform a radial
coordinate redefinition,
\be
r= \rho  -\ft12 m \sum_i s_i^2 + \fft{m^2}{3\rho}\, \sum_{i<j} 
            (s_i^2 - s_j^2)^2\,,
\ee
so that the components $h_{\mu\nu}$ fall off at large distance.
We find that the scalar potential energy and kinetic energy both diverge
(with linear and logarithmic divergences at large distance), but now, the
total scalar energy also diverges (with a linear, but no logarithmic, 
divergence):
\bea
&&-\fft{1}{8\pi \sqrt{-\bar g}}\, T^{t\nu}_{\rm pot}\, \bar\xi_\nu  = 
-\ft{1}{4\pi}g^2m^2 \sum_{i<j} (s_i^2 - s_j^2)^2\, \sin\theta\nn\\
&& +
\fft{3g^2 m^3 (s_1^2 + s_2^2 - s_3^2 - s_4^2)(s_1^2 - s_2^2 +
s_3^2 - s_4^2)(s_1^2 - s_2^2-s_3^2 +s_4^2)}{4\pi\, \rho}\, \sin\theta 
      + {\cal O}(1/\rho^2)\,,\nn\\
&&-\fft{1}{8\pi \sqrt{-\bar g}}\, T^{t\nu}_{\rm kinetic}\, \bar\xi_\nu =
\ft{1}{8\pi}g^2m^2 \sum_{i<j} (s_i^2 - s_j^2)^2\, \sin\theta     \\
&& -
\fft{3g^2 m^3 (s_1^2 + s_2^2 - s_3^2 - s_4^2)(s_1^2 - s_2^2 +
s_3^2 - s_4^2)(s_1^2 - s_2^2-s_3^2 +s_4^2)}{4\pi\, \rho}\, \sin\theta 
      + {\cal O}(1/\rho^2)\,.\nn
\eea

   To conclude this section, we shall consider two possible options for
modifying the AD prescription, so as to obtain the proper expressions
for the mass in the presence of scalar fields.  In fact, both have featured
in earlier discussions in the literature.
\bigskip

\noindent{\underline{\bf Option A:}}
\medskip

   The divergence in the volume integral for the total energy of the 
scalar fields in four dimensions
can in fact be removed, if one makes an integration by parts in the 
kinetic energy contribution for each scalar, of the form
\bea
-\fft1{8\pi}\, \int_{V} d^3x T^{t\nu}_{\rm kinetic}\,  \bar\xi_\nu &=& 
\fft1{32\pi}\, \int_V \sqrt{-\bar g}\, (\del\phi)^2\, \, d^3x \nn\\
&=&
 - \fft1{32\pi}\int_V \sqrt{-\bar g}\, \phi \bar{\square}\phi\, d^3 x
   + \fft1{32\pi}\, \int_{V} \del^\mu(\sqrt{-\bar g}\,\phi 
         \del^\mu \phi) \, d^3 x\nn\\
&=&  - \fft1{32\pi}\int_V \sqrt{-\bar g}\, \phi \bar{\square}\phi\, d^3 x
      + \fft1{32\pi}\, \oint_{\del V} dS_i\sqrt{-\bar g}\, 
          \phi \del^i\phi\,,
\eea
and then uses $-\phi\, \bar{\square}\phi$ rather than $(\del\phi)^2$ in the 
definition of the bulk energy-momentum tensor for each scalar field.  This
suggests therefore that one could define a ``corrected'' AD mass in 
situations where there are scalar fields, in which one adds an extra 
scalar boundary term to the expression given in (\ref{admass}), so that
\be
E = \fft1{8\pi}\, \oint dS_i ({\cal M}^i + {\cal N}^i)\,,\label{admodmass}
\ee
where ${\cal M}^i$ is still as given in (\ref{admass}), and the extra
term ${\cal N}^i$ is given by
\be
{\cal N}^i = -\ft14 \sqrt{-\bar g}\, \bar g^{ij}\, 
         {\cal G}_{IJ}(\phi)\, \phi^I\, \del_j\phi^J\,.\label{ndef}
\ee
Note that here, we are allowing for the general situation of scalar
fields $\phi^I$ with a non-linear sigma-model kinetic Lagrangian given by
\be
{\cal L}^{\rm kinetic} = -\ft12 {\cal G}_{IJ}(\phi)\, \del_\mu \phi^I\, 
       \del^\mu\phi^J\,. 
\ee
    
    Remarkably, we find that with the inclusion of the ${\cal N}^i$ term 
in the AD formula, we now obtain the correct results for the mass of the
3-charge black holes (\ref{3chargesol}) in five dimensions, and also for
the rotating black holes in five dimensions that were described in sections 
\ref{3chargebhsec}, \ref{1chargebhsec} and \ref{3chargeeqsec},
 and for the rotating black holes
that were described in section \ref{d4bhsec}, where we previously obtained 
incorrect masses using the AD prescription. It also leaves unaffected
the already-correct result in seven dimensions.  In fact we have only found 
one example where (\ref{admodmass}) with (\ref{ndef}) 
fails to give the correct result for
the mass, and that, ironically enough, is the very example that we used for
motivating the introduction of the correction term, namely the
four-dimensional non-rotating black hole with 4 unequal charges.  (If the 
charges are set pairwise equal, which corresponds to the non-rotating limit of
the rotating black-hole solution of \cite{d4gauge}, then (\ref{admodmass})
gives the correct mass.)  

      The extra term 
involving ${\cal N}^i$ that we have added to the AD calculation of the
mass is strikingly similar to the
surface term introduced by Hertog, Horowitz and Maeda in a discussion of
negative-energy solutions in five-dimensional maximal gauged supergravity
\cite{hehoma}.  There, a surface term of the form $\oint d S_\mu\, \phi
\, \del^\mu\phi$ was added to the action for a scalar field $\phi$, 
leading to a corresponding correction to the Hamiltonian and hence to the
mass.\footnote{We thank Gary Gibbons for drawing our attention to 
\cite{hehoma} after this paper was completed.} 

\bigskip
\noindent{\underline{\bf Option B:}}
\medskip

   A detailed discussion of the definition of mass in asymptotically AdS
backgrounds with scalar fields has been given in \cite{hematrza1,hematrza2}.
The focus in these papers was on cases where the scalars have masses that
saturate the Breitenl\"ohner-Freedman stability bound, namely
\be
m^2 = -\fft{(D-1)^2}{4 l^2}
\ee
in $D$ dimensions.  The reason for considering these limiting cases was 
that the possibility then arises of a less rapid fall-off for the scalar
fields fields, with a logarithmic dependence on the asymptotic radial
coordinate.  However, even in the absence of this logarithmic behaviour,
it was shown in \cite{hematrza1,hematrza2} that the scalar fields can
provide a contribution to the total energy, if one makes a decomposition
of the full metric as in (\ref{hdef}).  Translated into the notation
that we are using in this paper, we find that the scalar surface-integral
modifications of the type considered in \cite{hematrza2} can be expressed
as a different modification of the AD mass formula (\ref{admass}), 
analogous to the modification in (\ref{admodmass}), except that now we have
\be
E = \fft1{8\pi}\, \oint dS_i ({\cal M}^i + 
\widetilde {\cal N}^i)\,,\label{admodmass2}
\ee
with $\widetilde{\cal N}^i$ given by
\be
\widetilde{\cal N}^r= \fft{\sqrt{-\bar g}\, r}{4(D-1)}\, 
     \Big( \phi^I\, \phi^J\, \fft{\del^2 V(0)}{\del\phi^I\, \del\phi^J}
    + {\cal G}_{IJ}(\phi)\, g^{ij}\, \del_i\phi^I\, \del_j\phi^J\Big)\,.
\label{ndef2}
\ee
Note that the first term in the large parentheses is just the contribution
of the squared masses of the scalar fields, since a scalar field $\phi$ 
has mass squared given by
\be
m^2 = -\fft{\del^2 V(0)}{\del\phi^2}\,.
\ee

   A more compact way to write the correction term in this AD mass formula
is
\be
\widetilde{\cal N}^r = -\fft{ r}{D-1}\, (T_{\rm scal})^t{}_t\,,
\label{ndef3}
\ee
where $T^{\mu\nu}_{\rm scal}$ is the effective energy-momentum tensor 
density for the scalar fields, as defined in equation (\ref{scalart}).

   We find that using this modification, the AD mass formula 
(\ref{admodmass2}) with (\ref{ndef2}) or (\ref{ndef3}) 
then gives expressions that agree in all cases with
those that we obtained by using the AMD mass formula.  This includes
not only all the rotating black-hole solutions, but also the non-rotating
four-dimensional solution with four unequal charges, for which a
discrepancy still remained if we used the modification (\ref{admodmass}) 
with (\ref{ndef}).

   It is worth recording that except in five fimensions, the scalar
fields participating in the black-hole solutions in the various gauged
supergravities that we have been considering have (mass)$^2$ values
that exceed the Breitenl\"ohner-Freedman bound.  Specifically, the
relevant scalars arising in the theories in $D=4,5,6$ and 7 have
$m^2=(-2,-4,-6,-8)l^{-2}$, which can be contrasted with the
corresponding Breitenl\"ohner-Freedman masses $m^2=(-9/4, -4, -25/4,
-9)l^{-2}$. Of course in all cases, the scalar fields we are using
have apparent mass terms that imply ``masslessness'' in the
appropriate sense of being members of massless supermultiplets.
 
    Finally, it should be emphasised that if instead we use the AMD
procedure to calculate the masses for the general non-rotating black
holes in five, seven, six and four dimensions, and the various known
rotating black holes, we always get the correct mass without the need
for including any scalar modification terms.  In other words, the AMD
procedure yields the correct results for the masses by making reference
only to the metric. 

\section{Discussion and Conclusions}

   In this paper, we have principally been concerned with calculating the
masses of the various recently-discovered rotating black hole solutions 
in gauged supergravities in five, four and seven dimensions.  Until now,
the masses for the examples we have considered in this paper had been
calculated only by integrating the first law of thermodynamics.  This
has proved to be a reliable and straightforward procedure for calculating
the mass, which avoids some of the ambiguities inherent in certain other
approaches.  On the other hand, the thermodynamic calculation is somewhat
indirect, and does not emphasise the explicit relation between the energy and
a conservation law.  For this reason, it is of considerable interest to
perform alternative calculations of the masses of the rotating black
holes, that more directly relate the answer to conservation laws.

   We have focused on two such approaches in this paper, each of
which comes with its attendant advantages and disadvantages.  We first
considered the Ashtekar-Magnon-Das procedure for calculating the mass
of an asymptotically AdS spacetime.  The AMD procedure involves
integrating a certain electric component of the Weyl tensor over the
spatial section of the compactified conformal boundary.  It is
inherently well-defined, without the need for any subtraction, since
the Weyl tensor falls off suitably rapidly at large distance as the
metric approaches AdS spacetime.  The calculation is insensitive to
coordinate choices, and we have encountered no ambiguities at all when
calculating the masses for the various rotating black holes in gauged
supergravities that are known.  Furthermore, the masses that we have
calculated using the AMD prescription agree with the those obtained
previously by integrating the first law of thermodynamics.

  The only difficulties that one encounters when following the AMD
procedure are of a purely calculational nature, in that one has to
evaluate a certain component of the Weyl tensor of the full black-hole
metric (at least to leading order in an expansion in powers of the
inverse distance).  By contrast, the Abbott-Deser approach would be
computationally somewhat simpler, since one need only take first
derivatives of the deviation $h_{\mu\nu}$ and the timelike Killing vector, and
furthermore the derivatives are covariant just with respect to the
background AdS metric.  In practice the calculations for rotating
asymptotically AdS black holes are sufficiently complicated that in
either approach it is highly advantageous to use a computer, and so
the greater complexity of the AMD approach does not necessarily
represent a severe obstacle.

    In certain cases the AD approach is relatively easy to implement,
for example in the uncharged rotating AdS metrics as discussed in
\cite{dekate}, and in the general charged rotating solutions of
five-dimensional minimal gauged supergravity, which we analysed in
section \ref{d5minsec}.  The idea is to write the asymptotically AdS
metric $g_{\mu\nu}$ as the sum of a background AdS metric $\bar
g_{\mu\nu}$ plus deviation terms $h_{\mu\nu}$, and work in a frame
where $h_{\mu\nu}$ falls of appropriately at infinity.  However, we 
encountered difficulties when trying to apply the AD procedure to 
the five and four-dimensional rotating black hole solutions in which scalar
fields play a non-trivial r\^ole.  We considered two ways to modify the
original AD definition of the mass, to incorporate the effects of the
scalar fields.  One of these, given by (\ref{admodmass}) with (\ref{ndef}), 
is related to
a modification introduced in \cite{hehoma}.  We found that it then
led to agreement with the AMD and thermodynamic masses for all the
known rotating AdS black-holes solutions, but a discrepancy remained in
the case of four-dimensional AdS black holes with four unequal charges.
We then considered a different modification to the AD mass formula, 
generalising one introduced in \cite{hematrza2}, and we found that this,
given by (\ref{admodmass2}) with (\ref{ndef2}) or (\ref{ndef3}),
led to agreement with the AMD and thermodynamic masses for all the
black hole examples, including the four-dimensional black hole with
four unequal charges.

   Since the AMD approach gives reliable results for the masses of all
the black holes, which agree with the thermodynamic calculations
without the need for any scalar modifications, it would seem to be a
more ``robust'' prescription than the AD approach.  It yields expressions
for the masses by making reference only to the metric iteslf, and 
not to the scalar fields.  Furthermore, the
AMD approach does not involve the potentially hazardous process of
decomposing the black hole metric as a deviation $h_{\mu\nu}$ from a
background AdS metric $\bar g_{\mu\nu}$.  The hazards of this
decomposition are highlighted in the example of the non-rotating
four-dimensional gauged supergravity black hole with four unequal
charges, where the integral of the background-subtracted scalar energy
density is actually divergent at large distance.  This raises
questions about the validity of the assumption that the deviation
$h_{\mu\nu}$ is sufficiently small asymptotically.  More generally,
the whole question of how one should split the solution into
background plus deviation in the AD approach is somewhat unclear, and
the results might be affected by choices of field variable or
coordinate reparameterisations beyond those that we have considered.

    There are other methods that have also been used in order to
calculate the mass of asymptotically AdS spacetimes.  For example, as
we mentioned in the introduction, one could use a Komar integral,
although there are complications associated with the need to
regularise the divergent result by performing a background AdS
subtraction.  (However, see \cite{derkat2} for a recent discussion of
this approach, and its relation to the AMD method.)  Another approach
to calculating the mass of asymptotically AdS spacetimes is via the
holographic stress tensor, introduced in the context of the AdS/CFT
correspondence in string theory
\cite{henske,balkra,hasosk,awajoh,sken,papske}.  As was shown in
\cite{papske} for the uncharged rotating AdS black holes, this yields
bulk masses that are in agreement with those obtained previously from
the integration of the first law and from the AMD approach.  It also
yields Casimir contributions, which are not relevant in a classical
discussion of the energy of a black hole, but which do play a r\^ole
in the AdS/CFT correspondence and the map to the boundary theory.  As
was demonstrated explicitly in \cite{gibperpop3}, for the most natural
choice of conformal boundary metric the Casimir energy is a pure
constant, independent of the parameters of the black hole, and thus
its inclusion need not complicate the discussion of the
thermodynamics.

   Finally, we also carried out a check of the consistency of the 
quantum statistical relation (\ref{qsr}) for the general charged rotating 
black holes in minimal five-dimensional gauged supergravity.  This 
involved calculating the Euclidean action for the solution, and comparing it
with the thermodynamic potential.  This calculation also is rather subject to 
subtraction ambiguities, since one has to subtract the action of a pure AdS 
spacetime with the same boundary as that of the black-hole metric in order
to obtain a finite Euclidean action.  We did not carry through this 
procedure for the other, more complicated, examples of charged rotating
black holes, leaving this for future investigation.

\section*{Acknowledgments}

   We are grateful to Dick Arnowitt, Zhiwei Chong, Stanley Deser, 
Gary Gibbons and Malcolm Perry  for discussions and comments, and to Marc
Henneaux for correspondence after the appearance of the first version of
this paper.

\end{document}